\documentclass[preprint,floatfix] {revtex4} 
\newcommand{\rvec}{\mathrm {\mathbf {r}}} 
\newcommand{\pvec}{\mathrm {\mathbf {p}}}
\newcommand\tab[1][0.5cm]{\hspace*{#1}} 
\usepackage{graphicx}
\usepackage{subfigure}
\usepackage{xcolor}
\usepackage{amsmath}
\usepackage{enumerate}
\usepackage{longtable}
\usepackage{tabularx}
\usepackage{multirow}
\usepackage{amssymb}
\usepackage{color, soul}
\definecolor{darkblue}{rgb}{0,0,0.5}
\setulcolor{darkblue}

\begin{document}

\title{Information entropy and complexity measure in generalized Kratzer potential}

\author{Sangita Majumdar}
\author{Neetik Mukherjee}
\author{Amlan K.~Roy}
\altaffiliation{Corresponding author. Email: akroy@iiserkol.ac.in, akroy6k@gmail.com.}
\affiliation{Department of Chemical Sciences\\
Indian Institute of Science Education and Research (IISER) Kolkata, 
Mohanpur-741246, Nadia, WB, India}

\begin{abstract}
{\color{red}Shannon entropy ($S$), Fisher information ($I$) and a measure equivalent to Fisher-Shannon complexity $(C_{IS})$
  of a ro-vibrational state of diatomic molecules (O$_2$, O$_2^+$, NO, NO$^+$) with generalized Kratzer potential is analyzed.
 \emph{Exact} analytical expression of $I_{\rvec}$ is derived for the arbitrary state, whereas the same could be done for $I_{\pvec}$ 
with $\{n,\ell,m=0\}$ state. It is found that shifting from neutral to the cationic system, 
$I_{\rvec}$ increases while $S_{\rvec}$ decreases, consistent with the interpretation of a localization in the probability 
distribution. Additionally, this study reveals that $C_{IS}$ increases with the number of nodes in a system.}
\vspace{5mm}

\vspace{5mm}
{\bf Keywords:} diatomic molecule, generalized Kratzer potential, Fisher information, Shannon entropy, Fisher-Shannon 
complexity.  


\end{abstract}
\maketitle

\section{Introduction}
Ever since the early days of quantum mechanics, diatomic molecular potentials have received much attention due to their 
importance to describe intra-molecular and intermolecular interactions as well as atomic-pair correlations. Over the years, 
a large number of such potentials have been adopted in a multitude of physical/chemical problems. Since the literature is 
quite vast, here we mention a few prominent ones, such as generalized Morse, Mie, Kratzer-Fues, pseudoharmonic, non-central,
 deformed Rosen-Morse, generalized Woods-Saxon, P\"oschl-Teller potential etc \cite{akcay12}. They have fundamental relevance 
and utility in quantum description of natural phenomena, processes and systems, not only in the 3D 
world, but also in non-relativistic and relativistic D-dimensional physics. \textcolor{red}{This work focuses on generalized Kratzer 
(Kratzer-type) potential, which is a simple realistic zero-order model for describing the vibration-rotation 
motion of diatomic molecules. It has important properties like (i) correct asymptotic behaviour at $r=0$ and 
$r=\infty$ (ii) exactly solvable for a given state with an arbitrary angular momentum (iii) allows the system to dissociate, which 
is forbidden for a harmonic oscillator-based model. The mixed energy spectrum of this potential contains
both discrete and continuum parts, corresponding to bound and scattering states respectively. While bound-state wave functions 
have been widely used in studies related to molecular spectroscopy, latter states are important in the context of  
the photo-dissociation process, diatomic molecular scattering, radiative recombination in an atom-atom collision \cite{adelio02}, etc. Recently its
use as a universal potential for diatomic molecules has been reported in two excellent review articles \cite{hooydonk99, hooydonk00}. 
Apart from these, this has also been heavily studied in molecular, chemical and solid-state physics, due to its general feature of true 
interaction energy as well as interatomic, inter-molecular and dynamical properties \cite{kratzer20, fues26, oyewumi10}.}

In recent years, there has been a burgeoning activity in studies related to information-theoretic measures in quantum 
chemical systems. The delocalizing properties of electronic distribution that characterize the quantum states of these 
molecular potentials have been analyzed quite extensively by information theoretical tools such as Shannon entropy ($S$) 
\cite{birula75}, Fisher information ($I$) \cite{romera05}, R\'enyi entropy ($R$) \cite{sen12}, Tsallis entropy ($T$) 
\cite{tsallis88} and Onicescu information energy ($E$) \cite{sen12}, in both position ($r$) and momentum ($p$) spaces. 
Amongst these $S$ and $I$ describe a system in a complementary way. The former is a global measure of the distribution of 
density. \textcolor{red}{It is closely connected to the concept of 
entropy and disorder in thermodynamics. High values of $S_{\rvec}$ are associated with low values of $S_{\pvec}$ indicating a highly 
delocalized single-particle density distribution in $r$ space, and a localized one in $p$ space. It has a close relationship with 
quantities like kinetic energy, magnetic susceptibility and also has applications in the study of the dynamics of atoms and molecules.} 
The \textit{total} Shannon entropy ($S_{t}$), 
a sum of its $r$- and $p$-space counterparts, obeys the familiar entropic uncertainty relation: 
$S_{\rvec} + S_{\pvec}\geq d(1+\mathrm{ln}{\pi}$), where $d$ is associated dimension of the system.
This is a stronger version of Heisenberg uncertainty relation as it incorporates higher order moments in it \textcolor{red}{\cite{birula75}}.
On the other hand, $I$ is a cornerstone which signifies local \textcolor{red}{inhomogeneity} of single-particle density 
\emph{$\rho(\rvec)$} of a system under consideration. It is well-known that $I$ escalates with localization 
\textcolor{red}{as well as fluctuation} in probability density and diminishes in case of well spread-out 
densities. Of late, a spontaneity has been observed in the application of this information tool in various fields of 
physics and chemistry, mainly due to the fact that, the translationally invariant $I$ may be used as a quantitative 
measure of the spatial distribution of single-particle probability density of a many-particle system. It resembles the 
familiar Weisz\"acker kinetic energy functional in density functional theory of electronic systems \cite{chakraborty12}.
Numerous fundamental equations of physics and some conservation laws have also been derived using the property of $I$ 
as a basic variable of extreme physical information \cite{frieden04, nagy03}. 
Because of its versatile nature, it has potential application in exploring other different phenomena such as Pauli effect 
\cite{nagy06a, toranzo14}, polarizability, ionization potential, steric effect \cite{nagy07, esquivel11}, elementary 
chemical reaction \cite{rosa10}, bond formation \cite{nalewajski08}, etc. 

\textcolor{red}{Statistical complexity ($C_{LMC}$) \cite{lopezruiz95}}, another relevant concept, which arises due to 
breakdown of symmetry, is explicitly related to the aforementioned \textcolor{red}{fundamental} information theoretical tools. 
It represents a combined effect of two complementary 
quantities, offering a qualitative idea of the organization, structure and correlation in a system. It has finite value 
in a state lying between two limiting cases of \textcolor{red}{complete order (maximum distance from equilibrium) and maximum disorder 
(at equilibrium). The statistical measure of complexity is nothing but the product of the information content $(H)$ and 
concentration of spatial distribution $(D)$, and can be written as $C_{LMC}=H.D$, which was later criticized \cite{catalan02} and 
modified \cite{sanchez05} to the form of $C_{LMC}=D.e^{S}$ ($S$ quantifies the information of the system) in order to satisfy 
few conditions; such as reaching minimal values for both extremely ordered and disordered limits, invariance under scaling, 
translation and replication. Various definitions were put forth in literature. Some notable ones
include Shiner, Davidson and Landsberg (SDL) \cite{landsberg98,shiner99}, Fisher-Shannon 
($C_{IS}$) \cite{romera04,sen07,angulo08}, Cram\'er-Rao \cite{dehesa06a,angulo08,antolin09}, generalized R\'enyi-like 
\cite{calbet01,martin06,romera08} complexity, etc.} Amongst these, $C_{IS}$ corresponds to a measure which probes a system 
in terms of complementary global and local factors, and also satisfies certain desirable properties in complexity, 
such as, invariance under translations and re-scaling transformations, invariance under replication, 
near-continuity, etc.\cite{sen12}. This has remarkable applications in the study of the atomic shell structure, ionization processes 
\cite{sen07,angulo08,antolin09,angulo08a}, as well as in molecular properties like energy, ionization potential, hardness and dipole 
moment in the localization-delocalization plane showing chemically significant pattern \cite{esquivel10}, molecular reactivity 
studies \cite{welearegay14}. Few elementary chemical reactions such as hydrogenic-abstraction reaction \cite{esquivel11a}, 
identity \textcolor{red}{ S$_{N}^2$} exchange reaction \cite{molina-espiritu12}, and also concurrent phenomena occurring at the transition 
region \cite{esquivel12} of these reactions have been investigated through composite information-theoretic measures 
in conjugate spaces.

During the last few years, there has been a growing interest in studies on information theoretic and complexity 
measures in various model and real quantum systems. It is worthwhile mentioning a few selected works, \emph{viz.,} 
Morse \cite{dehesa06}, modified Yukawa and Hulth\'en \cite{patil07}, Dirac-delta \cite{bouvrie11}, Teitz-Wei 
diatomic molecular model \cite{falaye14}, squared tangent well \cite{dong14}, ring-shaped modified Kratzer \cite{yahya14},  
symmetric trigonometric Rosen-Morse \cite{nath14}, pseudoharmonic \cite{yahya15}, hyperbolic double well \cite{sun15}, 
infinite circular well \cite{song15}, ring-shaped Mie \cite{yahya14,falaye15}, P\"osch-Teller \cite{yahya16,dehesa06}, 
harmonic oscillator, generalized Morse \cite{onate16}, Eckart \cite{pooja16}, Frost-Muslin \cite{idiodi16},  
Killingbeck, Eckart Manning Rosen, \textcolor{red}{confined hydrogen atom \cite{majumdar17, mukherjee18, mukherjee18a}} etc. 
It may be noted that, a vast amount of elegant works have 
been published on eigenfunctions, eigenvalues and other properties of Schr\"odinger equation with these potentials, 
employing a variety of analytical/numerical methods, differing in their range of sophistication and accuracy. However, 
their information theoretical investigation is a rather recent {\color{red}development}. Moreover, whilst most of these works on $I$ 
were carried out numerically, analytical works have been rather limited. In \cite{romera05}, the authors derived 
expressions for {\color{red}Fisher information in $r$ $(I_{\rvec})$, and $p$ $(I_{\pvec})$ spaces} for an arbitrary quantum state of a 
single particle in a central potential in terms of radial expectation values $\{ \langle p^2 \rangle, \langle r^{-2} \rangle \} $ and 
$ \{ \langle r^2 \rangle, \langle p^{-2} \rangle \} $.

\textcolor{red}{While several papers have reported the energy spectrum (variety of approximate theoretical methods, as well as 
exact solution) of this potential, our present work focuses on the probability distribution through information measures such as 
$I, S, C_{IS}$, in both $r,p$ spaces in an arbitrary state, characterized by quantum numbers $n,\ell,m$ respectively.} Starting from exact 
$r$-space wave functions (as reported in \cite{oyewumi10}, through a factorization method), we first derive closed-form 
analytical expressions for $I_{\rvec}$ for any given $n,\ell,m$. It has been possible to obtain similar expressions for $I_{\pvec}$ 
as well, but only when $m=0$.  The $p$-space wave function is numerically obtained from the Fourier transform of the $r$-space 
counterpart. Simplified analytical formulas are derived also for $S_{\rvec}$; in this case, however, closed-form expressions 
are not possible; rather these are written in terms of some entropy integrals involving classical orthogonal polynomials.
Accurate results are presented for $I_{\rvec}, I_{\pvec}, S_{\rvec}, S_{\pvec}$ as well as $C_{IS}$, for representative 
states, in four selected diatomic molecules (two homo-nuclear, two hetero-nuclear) including two cations, namely O$_2$, 
O$_2^+$, NO, NO$^+$. Results are carefully monitored in both ground and excited states.   
Also, the system-dependence of scaling parameter $b$ in complexity measure is discussed. The article is organized as 
follows: Sec.~II gives a brief description of the theoretical method to find probability distribution and the desired 
information measures in a particular state. Then Sec.~III offers a thorough discussion of the results; finally,
a few concluding remarks are stated in Sec.~IV.

\section {Methodology}
The generalized Kratzer potential \cite{oyewumi10} may be represented in the following form,
\begin{equation}
v(r)=\frac{x}{r}+\frac{y}{r^2}+z.
\end{equation}
This is usually recast in following two alternative equivalent ways: 
\begin{enumerate}
\item
Kratzer-Fues potential, given by, 
\begin{equation}
v(r)=-D_0\left(\frac{2r_0}{r}-\frac{{r_0}^2}{r^2}\right), \ \ \  \ \mathrm{with} \ \ x=-2D_0r_0, \  y=D_0{r_0}^2, \  z=0, 
\end{equation}
where $D_0$ and $r_0$ represent the dissociation energy between two atoms and equilibrium intermolecular separation 
respectively. 
\item
Modified Kratzer or Mie potential, expressed as, 
\begin{equation}
v(r)=-D_0\left(\frac{2r_0}{r}-\frac{{r_0}^2}{r^2}\right)+D_0=D_0\left(\frac{r-r_0}{r}\right)^2,
 \mathrm{with} \ \ x=-2D_0r_0, \  y=D_0{r_0}^2, \  z=D_0.
\end{equation}
\end{enumerate}
Without any loss of generality, the time-independent non-relativistic wave function for a single particle in co-ordinate space
can be written as ($\rvec = \{ r, \Omega \}$), 
\begin{equation} 
\Psi_{n,\ell,m} (\rvec) = \psi_{n, \ell}(r)  \ Y_{\ell,m} (\Omega), 
\end{equation}
with $r$ and $\Omega$ denoting radial distance and solid angle respectively. Here $\psi_{n,\ell}(r)$ corresponds to the radial part 
and $Y_{\ell,m}(\Omega)$ the spherical harmonics of atomic state, determined by quantum numbers $(n,\ell,m)$. In what follows, 
atomic unit is employed unless otherwise mentioned and $\rvec, \pvec$ subscripts denote quantities in full $r$ and $p$ 
spaces (including angular part) respectively. 

Adopting the potential in Eq.~(3), the relevant radial Schr\"odinger equation \textcolor{red}{of the motion of a particle
 with reduced mass $\mu$ in a spherically symmetric potential $v(r)$ becomes,  
\begin{equation}
\left[-\frac{1}{2\mu} \ \frac{d^2}{dr^2} + \frac{\ell (\ell+1)} {2r^2} + v(r) \right] \psi_{n,\ell}(r)=
{E}_{n,\ell}\ \psi_{n,\ell}(r).
\end{equation}}
Note that, throughout the present work, we shall deal with this form of potential, as Eq.~(2) constitutes a special case
of it. The angular part has following common form in $r, p$ spaces,  
\begin{equation}
Y_{\ell,m} ({\Omega}) =\Theta_{\ell,m}(\theta) \ \Phi_m (\phi) = (-1)^{m} \sqrt{\frac{2l+1}{4\pi}\frac{(\ell-m)!}{(\ell+m)!}} 
\ P_{\ell}^{m}(\cos \theta)\ e^{-im \phi}, 
\end{equation}
with $P_{\ell}^{m} (\cos \theta)$ signifying the usual associated Legendre polynomial.  
The exact normalized radial wave function for this potential has been obtained with the factorization method \cite{oyewumi10},
\begin{equation}
\psi_{n,\ell}(r) = N_{n,\ell} \ e^{-\frac{\xi r}{2}} r^{\beta_{\ell}} \ L^{2 {\beta_{\ell}} + 1}_{n} \left(\xi r \right)
\end{equation}
where \textcolor{red}{$L^{2 {\beta_{\ell}} + 1}_{n} \left(\xi r \right)$ is denoted the Associated Laguerre polynomial}
 and the normalization constant has form,
\begin{equation}
N_{n,\ell} = \left[\frac{\xi^{2 \beta_{\ell} + 3}}{2} \frac{n!}{\left(n + \beta_{\ell} +1\right) 
\Gamma \left(n + 2 \beta_{\ell} +2\right)}\right]^\frac{1}{2}.
\end{equation}
Here, the quantities $\xi$ and $\beta_{\ell}$ are defined as below, 
\begin{equation}
\xi = {\left[-8 \mu \left( {E}_{n,\ell} - z \right)\right]}^\frac{1}{2}, \ \ \ \ \ \
\beta_{\ell} = \frac{1}{2} \left[-1 + \sqrt{(2 \ell + 1)^2 + 8 \mu y}\right].
\end{equation}
The corresponding bound-state eigenvalues are written as, 
\begin{equation}
E_{n,\ell} = \frac{- \mu x^2}{2 \left[n + \beta_{\ell} + 1\right]^2} + z.
\end{equation}
The wave function in $\emph{p}$ space is obtained by taking usual Fourier transformation of the $\emph{r}$-space counterpart 
and as such given by the standard expression,  
\begin{equation}
\begin{aligned}
\Xi_{n,\ell}(p) & = & \frac{1}{(2\pi)^{\frac{3}{2}}} \  \int_0^\infty \int_0^\pi \int_0^{2\pi} \psi_{n,\ell}(r) \ \Theta(\theta) 
 \Phi(\phi) \ e^{ipr \cos \theta}  r^2 \sin \theta \ \mathrm{d}r \mathrm{d} \theta \mathrm{d} \phi ,       
\end{aligned}
\end{equation}
where $\Xi_{n,\ell}(p)$ needs to be normalized. \textcolor{red}{Integrations over $\theta$ and $\phi$ have been done 
analytically. Depending on $\ell$ the coefficients of integration vary; these have been discussed and tabulated in 
\cite{mukherjee18a}.}
Thus the normalized $r$- and $p$-space electron densities can be expressed as, 
$\rho(\rvec)=|\Psi_{n,\ell,m} (\rvec)|^2= |\psi_{n,\ell}(r)|^2 \ |Y_{\ell,m} (\Omega)|^2 $ and 
\textcolor{red}{$\Pi(\pvec)=|\Lambda_{n,\ell,m}(\pvec)|^2=|\Xi_{n,\ell}(p)|^2 \ |Y_{\ell,m} (\Omega)|^2 $} respectively.

Now, following \cite{romera05}, {\color{red}Fisher information} of a single particle in a central potential can be simplified in 
terms of radial expectation values in $r$ and $p$ spaces, as below {\color{red}\cite{romera05}}, 
\begin{equation} 
\begin{aligned} 
I_{\rvec}  & =  \int_{{\mathcal{R}}^3} \left[\frac{|\nabla\rho(\rvec)|^2}{\rho(\rvec)}\right] \mathrm{d}\rvec  =  
4\langle p^2\rangle - 2(2\ell+1)|m|\langle r^{-2}\rangle \\  
I_{\pvec} & =  \int_{{\mathcal{R}}^3} \left[\frac{|\nabla\Pi(\pvec)|^2}{\Pi(\pvec)}\right] \mathrm{d} \pvec  = 
4\langle r^2\rangle - 2(2\ell+1)|m|\langle p^{-2}\rangle.
\end{aligned} 
\end{equation}
When $m=0$, $I_{\rvec}$ and $I_{\pvec}$ assume simpler expressions,  
\begin{equation}
I_{\rvec}  =  4\langle p^2\rangle, \hspace{3mm} \ \ \ \  I_{\pvec} =4\langle r^2\rangle.
\end{equation}
It has been established that the general Heisenberg-like uncertainty relation associated with a particle in a $D$-dimensional 
central potential can be expressed as \cite{moreno06},
\begin{equation}
\left \langle r^{2}\right\rangle \left \langle p^{2}\right\rangle \ \geqslant \left(L + \frac{3}{2}\right)^2, 
\end{equation}
where $L = \ell + \frac{D-3}{2}$ is generalized angular momentum (grand orbital quantum number).
From this modified Heisenberg relation, $I$-based uncertainty relation transforms into \cite{dehesa07}, 
\begin{equation}
I_{\rvec}I_{\pvec} \geqslant 16\left(1 - \frac{2 |m|}{2L + 1}\right)^2 \left(L + \frac{3}{2}\right)^2.
\end{equation}
For the particular case of $D=3$, Eqs.~(14) and (15) lead to following bounds,  
\begin{equation}
\begin{aligned} 
\left \langle r^{2}\right\rangle \left \langle p^{2}\right\rangle  &  \geqslant \left(\ell + \frac{3}{2}\right)^2 \\
I_{\rvec}I_{\pvec} & \geqslant 16\left(1 - \frac{2 |m|}{2l + 1}\right)^2 \left(\ell + \frac{3}{2}\right)^2.
\end{aligned} 
\end{equation}

\begingroup 
\squeezetable
\begin{table}   
\caption{Spectroscopic parameters of diatomic molecules studied in this work, taken from \cite{falaye14}.} 
\centering
\begin{ruledtabular} 
\begin{tabular}{l|llll}
Molecule (state)       & $\mu/10^{-23}$ (g)    & $D_0$ (cm$^{-1})$  & $r_{0}$ (\AA)    \\ 
\hline
O$_{2}$ $(X^3\Sigma_{g}^{+})$     & 1.337        &  42041       &  1.207      \\ 
O$_{2}^{+}$ $(X^2\Pi_{g})$        & 1.337        &  54688       &  1.116      \\ 
NO $(X^2\Sigma_{r})$              & 1.249        &  53341       &  1.151      \\ 
NO$^{+}$ $(X^1\Sigma^{+})$        & 1.239        &  88694       &  1.063      \\ 
\end{tabular}
\end{ruledtabular} 
\end{table}
\endgroup 

Next, $S$ in conjugate $r$ and $p$ spaces (in spherical polar coordinates) can be written as, 
\begin{eqnarray}
\begin{aligned} 
S_{\rvec}  =  -\int_{{\mathcal{R}}^3} \rho(\rvec) \ \ln [\rho(\rvec)] \ \mathrm{d} \rvec   = 
2\pi \left(S_{r}+S_{(\theta,\phi)}\right) ,\\    
S_{\pvec}  =  -\int_{{\mathcal{R}}^3} \Pi(\pvec) \ \ln [\Pi(\pvec)] \ \mathrm{d} \pvec   = 
2\pi \left(S_{p}+S_{(\theta, \phi)}\right), \\
{\color{red}S_{\rvec}+S_{\pvec} =S_{t} \geq d(1+ \ln \pi),} 
\end{aligned} 
\end{eqnarray}
{\color{red}where d is the dimension of the system.} \\
The quantities $S_r, S_p$ and $S_{\theta}$ have been defined as \cite{birula75},   
\begin{eqnarray}
\begin{aligned} 
S_{r} & =  -\int_0^\infty \rho(r) \ \ln [\rho(r)] r^2 \mathrm{d}r,\ \hspace{25mm}  \rho(r ) = |\psi_{n,\ell}(r)|^{2}, \\ 
S_{p} & =  -\int_{0}^\infty \Pi(p) \ln [\Pi(p)]  \ p^2 \mathrm{d}p,   \hspace{22mm} {\color{red}\Pi(p)  = |\Xi_{n,\ell}(p)|^{2},} \\
S_{(\theta, \phi)} & =   -\int_0^\pi \chi(\theta) \ \ln [\chi(\theta)] \sin \theta \mathrm{d} \theta, \hspace{20mm} 
 \chi(\theta)  =  |\Theta(\theta)|^2.  
\end{aligned} 
\end{eqnarray}
Finally, the Fisher-Shannon complexity is defined as ($b$ is a scaling factor) \textcolor{red}{\cite{romera04}},
\begin{equation}
C_{IS}  = I e^{bS}.
\end{equation}

\section{Result and Discussion}
At first, it may be prudent to mention a few things to facilitate the forthcoming discussion. As hinted before, apart from 
$I_{\rvec}$ (all states) and $I_{\pvec}$ (only $m=0$), all other quantities like $I_{\pvec}$ ($m \neq 0$), as well as 
$S_{\rvec}$, $S_{\pvec}$ and $C_{IS}$, presented in this work are calculated numerically. All the tables and figures contain the 
\textit{net} information measures in conjugate $r$ and $p$ spaces, which can be separated into \textit{radial} and \textit{angular} 
components. It is evident from Eq.~(12) that in both $I_{\rvec}$ and $I_{\pvec}$ expressions, angular parts are normalized to unity; 
thus evaluation of these quantities using only radial parts will suffice the purpose. The complexity measures of Eq.~(19) have been 
probed for two selected values of $b$, \emph{viz.,} 1 and $\frac{2}{3}$, which are the most often used values in literature. 
Amongst these, the latter is usually referred as $C_{IS}$. \textcolor{red}{A simplified notation, $C_{I_{s},S_{s}}^{b}$ 
is used throughout our discussion for convenience, where the subscripts ``s" is used to specify the conjugate space 
\textit{r} or \textit{p}.} The superscript ``b" 
takes two values identifying two scaling parameters $\frac{2}{3}$ and $1$. All results are presented here for 
four representative diatomic molecules, namely, O$_2$, O$_2^+$, NO, NO$^+$, which have the model parameters listed in Table~I; 
these are adopted from reference \cite{falaye14}. It is worthwhile mentioning that, convergence of all numerically calculated 
quantities were carefully checked with respect to the grid parameters and some of these have been detailed earlier \cite{mukherjee18} 
in the context of a confined H atom embedded inside an impenetrable spherical cavity; hence not repeated here. These are 
checked for convergence up to the place 
they are given. Lastly, since the exact solution including energy spectrum as well as eigenfunctions have already been reported
by a number of authors in literature, we do not discuss these here any more.  

\begingroup            
\squeezetable
\begin{table}
\tiny
\caption{$I_\rvec^{\updownarrow\updownarrow},~I_\pvec^{\dagger\dagger}$ and {\color{red}$I_{t}$} for some selected states of four diatomic molecules. See text for details.}
\centering
\begin{ruledtabular} 
\begin{tabular}{l|lll|lll|lll|lll|c }
 & \multicolumn{3}{c|}{O$_{2}$} & \multicolumn{3}{c|}{O$_{2}^{+}$} & \multicolumn{3}{c|}{NO} & \multicolumn{3}{c|}{NO$^{+}$} &  \\
\hline
$n^{\updownarrow}$ & $I_{\rvec}$ & $I_{\pvec}$ &$I_{t}$ & $I_{\rvec}$ & $I_{\pvec}$ &$I_{t}$ & $I_{\rvec}$ & $I_{\pvec}$ &$I_{t}$ & $I_{\rvec}$ & $I_{\pvec}$ &$I_{t}$ & 
$I_{t}^{\P}$ \\
\hline
0   & 65.367653   & 21.239249 &{\color{red}1388.35 } &  80.657917  & 18.138248& {\color{red}1462.99 } &  74.644253  &  19.299666 &{\color{red}1440.60 } &   104.053919  &   16.410317 & {\color{red}1707.55} & 36 \\ 
1  &  192.338431  & 22.110305 &{\color{red}4252.66 } & 237.565501  & 18.843575& {\color{red}4476.58 } & 219.790637  &  20.062002 &{\color{red}4409.44 } &   307.288984  &   16.955810 & {\color{red}5210.33} & 36  \\
2  &  314.931505  & 23.000945 &{\color{red}7243.72 } & 389.338510  & 19.563941& {\color{red}7616.99 } & 360.113455  &  20.840849 &{\color{red}7505.07 } &   504.823940  &   17.511254 & {\color{red}8840.10} & 36  \\
3  &  433.296652  & 23.911420 &{\color{red}10360.73} & 536.143741  & 20.299530& {\color{red}10883.46} & 495.772076  &  21.636412 &{\color{red}10726.72} &   696.817662  &   18.076752 & {\color{red}12596.20} & 36  \\
4  &  547.577993  & 24.841985 &{\color{red}13602.92} & 678.142011  & 21.050528& {\color{red}14275.24} & 626.919800  &  22.448896 &{\color{red}14073.65} &   883.424127  &   18.652408 & {\color{red}16477.48} & 36  \\
5  &  657.914225  & 25.792896 &{\color{red}16969.51} & 815.488386  & 21.817121& {\color{red}17791.60} & 753.704368  &  23.278510 &{\color{red}17545.11} &  1064.792586  &   19.238329 & {\color{red}20484.83} & 36  \\
\hline
$\ell^{\dagger}$ & \\
\hline
 0 & 657.914225  & 25.792896 & {\color{red}16969.51} & 815.488386  & 21.817121 & {\color{red}17791.60} & 753.704368  &  23.27851  &{\color{red}17545.11} &  1064.792586  &   19.238329 &{\color{red}20484.83} & 36   \\ 
 1 & 659.247075  & 25.796242 & {\color{red}17006.09} & 817.058906  & 21.819672 & {\color{red}17827.95} & 755.17771   &  23.281317 &{\color{red}17581.53} &  1066.557661  &   19.239985 &{\color{red}20520.55}  & 100   \\
 2 & 661.911687  & 25.802935 & {\color{red}17079.26} & 820.198793  & 21.824776 & {\color{red}17900.65} & 758.123278  &  23.286931 &{\color{red}17654.36} &  1070.086862  &   19.243298 &{\color{red}20592.00}  & 196   \\
 3 & 665.905888  & 25.812976 & {\color{red}17189.01} & 824.905745  & 21.832433 & {\color{red}18009.69} & 762.538843  &  23.295354 &{\color{red}17763.61} &  1075.378295  &   19.248269 &{\color{red}20699.17}  & 324   \\
 4 & 671.226423  & 25.826367 & {\color{red}17335.33} & 831.176311  & 21.842644 & {\color{red}18155.08} & 768.421066  &  23.306587 &{\color{red}17909.27} &  1082.429122  &   19.254896 &{\color{red}20842.06}   & 484   \\
 5 & 677.868959  & 25.843111 & {\color{red}17518.24} & 839.005895  & 21.855411 & {\color{red}18336.81} & 775.7655    &  23.320632 &{\color{red}18091.34} &  1091.235561  &   19.263183 &{\color{red}21020.67}  & 676   \\
\hline
$m^{\ddagger}$ & \\
\hline
 0   &  677.868959 & 25.843111 &{\color{red}17518.24} & 839.005895  & 21.855411 & {\color{red}18336.81} & 775.765500  &  23.320632  &{\color{red}18091.34} & 1091.235561  &   19.263183  &{\color{red}21020.67} & 676    \\ 
 1   &  674.031036 & 25.249005 &{\color{red}17018.61} & 834.493874  & 21.369094 & {\color{red}17832.37} & 771.529881  &  22.7966953 &{\color{red}17588.33} & 1086.195415  &   18.8743977 &{\color{red}20501.28} &  $\frac{54756}{121}$ \\
 2   &  670.193113 & 24.654900 &{\color{red}16523.54} & 829.981852  & 20.882778 & {\color{red}17332.32} & 767.294262  &  22.2727585 &{\color{red}17089.75} & 1081.155269  &   18.4856120 &{\color{red}19985.81} &  $\frac{33124}{121}$ \\
 3   &  666.355189 & 24.060795 &{\color{red}16033.03} & 825.469831  & 20.396461 & {\color{red}16836.66} & 763.058644  &  21.7488218 &{\color{red}16595.62} & 1076.115123  &   18.096826  &{\color{red}19474.26} &  $\frac{16900}{121}$ \\
 4   &  662.517266 & 23.466689 &{\color{red}15547.08} & 820.957809  & 19.910144 & {\color{red}16345.38} & 758.823025  &  21.2248850 &{\color{red}16105.93} & 1071.074976  &   17.708040  &{\color{red}18966.63} &  $\frac{6084}{121} $ \\
 5   &  658.679343 & 22.872584 &{\color{red}15065.69} & 816.445788  & 19.423827 & {\color{red}15858.50} & 754.587406  &  20.7009483 &{\color{red}15620.67} & 1066.03483   &   17.31925   &{\color{red}18462.92} &  $\frac{676}{121}  $ \\
\end{tabular}
\end{ruledtabular}
\begin{tabbing}
\textcolor{red}{$^{\updownarrow}\ell$ and $m$ are fixed at $0$.  \tab   $^{\dagger}n$ and $m$ are fixed at $5$ and $0$ respectively. 
\tab $^{\ddagger}n$ and $\ell$ both are fixed at $5$.}  \tab   
$^{\updownarrow\updownarrow}$Calculated from Eq.~(27). \\ 
$^{\dagger\dagger}$In a state having $m=0$, these are calculated from Eq.~(28). For all other states, these are obtained 
numerically.   \\ 
$^{\P}$Lower bounds of $I_{t}=I_{\rvec}I_{\pvec}$, obtained from Eq.~(16). In all cases, the products satisfy this bound.       
\end{tabbing}
\end{table}
\endgroup

Now let us proceed for evaluation of $I_{\rvec}$ and $I_{\pvec}$, for which some expectation values are needed. This 
can be achieved by means of the following sequence of steps; first is to write the Hamiltonian in following convenient form, 
\begin{equation}
H = \frac{p^2}{2 \mu} + \left[-D_0\left(\frac{2r_0}{r} - \frac{{r_0}^2}{r^2}\right) + z\right].
\end{equation}
This prompts us to write $\left\langle p^{2}\right\rangle$ as, 
\begin{equation}
\langle p^2 \rangle = 2 \mu \langle H \rangle - 2 \mu z + 4 \mu D_0 r_0 \left\langle \frac{1}{r} \right\rangle -
2 \mu D_0 r_{0}^2 \left\langle \frac{1}{r^2} \right\rangle, 
\end{equation}
from which it is apparent that, in order to evaluate $\left\langle p^{2}\right\rangle$, one is required to compute the additional 
expectation values such as: $\left\langle H\right\rangle$, $\left\langle \frac{1}{r}\right\rangle$ and 
$\left\langle \frac{1}{r^{2}}\right\rangle$. The first term $\left\langle H\right\rangle$ is nothing but the energy 
eigenvalue of a particular state with definite values of $n,\ell$, and can be easily calculated from Eq.~(10).
Next, we turn to $\left\langle \frac{1}{r}\right\rangle$, which can be carried out as below,
\begin{equation}
\left\langle \frac{1}{r}\right\rangle = \int_0^\infty |N_{n,\ell}|^2  e^{{\xi} r} r^{2 \beta_{\ell} + 1} 
\left[L^{2 {\beta_{\ell}} + 1}_{n}(r)\right]^2  \mathrm{d}r  
 = |N_{n,\ell}|^2 \left(\frac{1}{\xi}\right)^{\left(2 \beta_{\ell} +2\right)} 
\frac{\left(n + 2 \beta_{\ell} + 1\right)!}{n!},
\end{equation}
where we have used the standard integral form of Associated Laguerre polynomial \cite{gradshteyn07},
\begin{equation} 
\int_0^\infty t^{\alpha} e^{-t} \left[L^{\alpha}_{n}(t)\right]^2 \mathrm{d}t = \frac{(n+\alpha)!}{n!} .
\end{equation}
Also, $\left\langle \frac{1}{r^{2}}\right\rangle$ can be calculated as,
\begin{equation}
\begin{aligned}
\left\langle \frac{1}{r^2}\right\rangle &= \int_0^\infty |N_{n,\ell}|^2  e^{{\xi} r} r^{2 \beta_{\ell}}
\left[L^{2 {\beta_{\ell}} + 1}_{n}(r)\right]^2  \mathrm{d}r  \\
&= |N_{n,\ell}|^2 \left(\frac{1}{\xi}\right)^{\left(2 \beta_{\ell} +1\right)} \sum_{i=0}^n \left(\begin{array}{c}
 -1\\ n-i \end{array}\right)^2 \frac{\Gamma\left(2 \beta_{\ell} +1 + i\right)}{i!} 
\end{aligned}
\end{equation}
which makes use of the following standard integral \cite{gradshteyn07},
\begin{equation} 
\int_0^\infty t^{\alpha + \beta} e^{-t} \left[L^{\alpha}_{n}(t)
 \right]^2 \mathrm{d}t = \sum_{i=0}^n \left(\begin{array}{c} \beta\\ n-i \end{array}\right)^2 
 \frac{\Gamma\left(\alpha + \beta +1 + i\right)}{i!}.
\end{equation}
Finally, $\left\langle r^{2} \right\rangle$ reads,
\begin{equation}
\begin{aligned}
\left\langle {r^2}\right\rangle & = \int_0^\infty |N_{n,\ell}|^2  e^{{\xi} r} r^{2 \beta_{\ell}+4}
\left[L^{2 {\beta_{\ell}} + 1}_{n}(r)\right]^2  \mathrm{d}r  \\
&= |N_{n,\ell}|^2 \left(\frac{1}{\xi}\right)^{\left(2 \beta_{\ell} +5\right)} \sum_{i=0}^n \left(\begin{array}{c}
 3\\ n-i \end{array}\right)^2 \frac{\Gamma\left(2 \beta_{\ell} +5 + i\right)}{i!}, 
\end{aligned}
\end{equation}
where once again the standard integral form given in Eq.~(25) has been utilized. 

Next the substitution of radial expectations values $\left\langle p^{2} \right\rangle$ and $\left\langle r^{-2} \right\rangle$, 
of Eqs.~(21) and (24) in Eq.~(12) yields the desired expression of $I_{\rvec}$ in a particular state,  
\begin{equation}
\begin{aligned}
I_{\rvec} & = 4\left[2 \mu \left \langle H\right \rangle - 2\mu z + 4 \mu D_{0} r_{0} \left\langle \frac{1}{r} \right\rangle 
- 2 \mu D_{0}
{r_{0}}^2 \left\langle \frac{1}{r^2}\right\rangle  \right] - 2(2 \ell + 1) |m| \left\langle \frac{1}{r^{2}} \right\rangle \\
&= 8 \mu E_{n, \ell} - 8 \mu z + 16 \mu D_{0} r_{0} \left[ |N_{n,\ell}|^2 \left(\frac{1}{\xi}\right)^{\left(2 \beta_{\ell} 
+2\right)} \frac{\left(n + 2 \beta_{\ell} + 1\right)!}{n!} \right] - \\ 
&\left\{ 8 \mu D_{0} {r_{0}}^{2} - 2 (2 \ell + 1) |m| \right \} \left[ |N_{n,\ell}|^2 \left(\frac{1}{\xi}\right)^
{\left(2 \beta_{\ell} +1\right)} \sum_{i=0}^n \left(\begin{array}{c} -1\\ n-i \end{array}\right)^2 
\frac{\Gamma\left(2 \beta_{\ell} +1 + i\right)}{i!} \right].
\end{aligned}
\end{equation}
Similarly, $I_{\pvec}$ for $m=0$ states, can be simplified as, 
\begin{equation}
I_{\pvec} = 4 \left \langle r^{2} \right \rangle 
= 4 |N_{n, \ell}|^{2} \left(\frac{1}{\xi}\right)^{\left(2 \beta_{\ell} +5\right)} \sum_{i=0}^n \left(\begin{array}{c}
 3\\ n-i \end{array}\right)^2 \frac{\Gamma\left(2 \beta_{\ell} +5 + i\right)}{i!},  
\end{equation}
where the expression of $\langle r^2 \rangle$ in Eq.~(26) has been employed.

Now we are ready to present our estimated $I_{\rvec}$ and $I_{\pvec}$ values in Table~II. Towards this goal, a cross-section 
of results are given for O$_{2}$, O$_{2}^{+}$, NO and NO$^{+}$. For each 
molecule, these dual measures in conjugate spaces are given in three horizontally separated regions; each one referring to 
the variation of one state index, keeping other two fixed. Thus the top, middle and bottom segments characterize states
varying $n, \ell$ and $m$ respectively. State indices alter in the range of $0-5$. 
It is noticed that, in all these states, both $I_{\rvec}, I_{\pvec}$ increase with an increase in $n$ for fixed $\ell,m$. This is to 
be expected, as, with rise in $n$, number of nodes grows, which promotes fluctuation. Equation~(12) suggests that for $m=0$ states, 
$I_{\rvec}$ may be associated 
with the kinetic energy of the system. Hence, as $n$ (at fixed $\ell,m$) advances, $I_{\rvec}$ accumulates indicating a rise in 
kinetic energy. The table also reflects that, both $I_{\rvec}, I_{\pvec}$ progress with $\ell$ (at fixed $n,m$). Note that, in 
this potential, the number of nodes in a given state depends only on $n$. Therefore, at some particular 
$n,m$ values, the rise in $I_{\rvec}$ and $I_{\pvec}$  with $\ell$ indicates the enhancement of fluctuation in states having 
a fixed number of nodes. However, they both decline with growth in $m$ (at fixed $n,\ell$). 
Furthermore, $I_{\rvec}$ in cationic systems possess higher values than their neutral counterparts. 
\textcolor{red}{In both O$_2$ and NO, on going from the neutral to cationic species, $D_{0}$ increases while $r_{0}$ decreases,
 indicating an increase in bond strength. This results in an enhancement of localization, which is reflected in the growth of 
$I_{\rvec}$.}
In addition, from the reported values in this table, $I_{t}$ can easily be found to satisfy the lower 
bound given in Eq. (16) in terms of $I_{\rvec}$ and $I_{\pvec}$.

\begingroup            
\squeezetable
\begin{table}
\tiny
\caption{$S_\rvec,~S_\pvec$ and {\color{red}$S_{t}^{\P}$} for some selected states of four diatomic molecules. See text for details.}
\centering
\begin{ruledtabular} 
\begin{tabular}{ l|lll|lll|lll|lll }
  &  \multicolumn{3}{c|}{O$_{2}$} & \multicolumn{3}{c|}{O$_{2}^{+}$} & \multicolumn{3}{c|}{NO} & \multicolumn{3}{c}{NO$^{+}$} \\
\hline
$n^{\updownarrow}$ & $S_{\rvec}$ & $S_{\pvec}$ &$S_{t}$& $S_{\rvec}$ & $S_{\pvec}$ &$S_{t}$& $S_{\rvec}$ & $S_{\pvec}$ & $S_{t}$&$S_{\rvec}$ & $S_{\pvec}$&$S_{t}$ \\
\hline
0 &   3.5256409472  & 6.0023 & {\color{red}9.5279}  & 3.2629165502 & 6.3142  &{\color{red} 9.5771} & 3.3636640774  &  6.1989  &{\color{red} 9.5625} & 3.0359854245  &   6.6868 &{\color{red}9.7227}   \\ 
1 &   3.831160      & 8.1101 & {\color{red}11.9412} & 3.566630     & 8.4270  &{\color{red}11.9936} & 3.667899      &  8.3103  &{\color{red}11.9781} & 3.334892      &   8.8133 &{\color{red}12.1481}   \\
2 &   4.021609      & 8.3220 & {\color{red}12.3436} & 3.755314     & 8.6378  &{\color{red}12.3931} & 3.857093      &  8.5214  &{\color{red}12.3784} & 3.518875      &   9.0207 &{\color{red}12.5395}  \\
3 &   4.166878      & 9.0066 & {\color{red}13.1734} & 3.898848     & 9.3257  &{\color{red}13.2245} & 4.001129      &  9.2084  &{\color{red}13.2095} & 3.65778       &   9.7173 &{\color{red}13.3750}  \\
4 &   4.287550      & 9.1319 & {\color{red}13.4194} & 4.017814     & 9.4500  &{\color{red}13.4678} & 4.120588      &  9.3330  &{\color{red}13.4535} & 3.77219       &   9.8398 &{\color{red}13.6119}  \\
5 &   4.392598      & 9.5278 & {\color{red}13.9203} & 4.121181     & 9.8486  &{\color{red}13.9697} & 4.224441      &  9.7308  &{\color{red}13.9552} & 3.87107       &   10.2464&{\color{red}14.1174}   \\
\hline 
$\ell^{\dagger}$ & \\
\hline      
0 &  4.392598  & 9.5278 &{\color{red}13.9203} & 4.121181  & 9.8486 &{\color{red}13.9697} & 4.224441  &  9.7308 &{\color{red}13.9552} &  3.87107  &   10.2464 &{\color{red}14.1174} \\ 
1 &  3.960834  & 9.109  &{\color{red}13.0698} & 3.689399  & 9.4294 &{\color{red}13.1187} & 3.792664  &  9.3117 &{\color{red}13.1043} &  3.43924  &   9.8241  &{\color{red}13.2633} \\
2 &  3.903244  & 9.0711 &{\color{red}12.9743} & 3.631773  & 9.3903 &{\color{red}13.0220} & 3.735048  &  9.273  &{\color{red}13.0080} &  3.38153  &   9.7835  &{\color{red}13.1200} \\
3 &  3.883322  & 9.0749 &{\color{red}12.9582} & 3.611798  & 9.3934 &{\color{red}13.0051} & 3.715089  &  9.2763 &{\color{red}12.9913} &  3.36142  &   9.7832  &{\color{red}13.1446} \\
4 &  3.873926  & 9.0938 &{\color{red}12.9677} & 3.602330  & 9.4105 &{\color{red}13.0128} & 3.705641  &  9.2938 &{\color{red}12.9994} &  3.35178  &   9.7965  &{\color{red}13.1482} \\
5 &  3.868874  & 9.1189 &{\color{red}12.9877} & 3.597188  & 9.4348 &{\color{red}13.0319} & 3.700524  &  9.3185 &{\color{red}13.0190} &  3.34642  &   9.8174  &{\color{red}13.1638} \\
\hline 
$m^{\ddagger}$ & \\
\hline      
 0   &  3.868874   & 9.1189 &{\color{red}12.9877} & 3.597188  & 9.4348 &{\color{red}13.0319} & 3.700524  &  9.3185 &{\color{red}13.0190} &  3.34642  &   9.8174 &{\color{red}13.1638} \\ 
 1   &  3.9858     & 9.2358 &{\color{red}13.2216} & 3.714114  & 9.5517 &{\color{red}13.2658} & 3.81745   &  9.4354 &{\color{red}13.2528} &  3.46335  &   9.9344 &{\color{red}13.3977} \\
 2   &  4.040809   & 9.2908 &{\color{red}13.3316} & 3.769123  & 9.6067 &{\color{red}13.3758} & 3.872459  &  9.4904 &{\color{red}13.3628} &  3.51836  &   9.9894 &{\color{red}13.5077} \\
 3   &  4.049226   & 9.2992 &{\color{red}13.3484} & 3.77754   & 9.6152 &{\color{red}13.3927} & 3.880876  &  9.4988 &{\color{red}13.3796} &  3.52677  &   9.9978 &{\color{red}13.5245} \\
 4   &  4.000479   & 9.2505 &{\color{red}13.2509} & 3.728793  & 9.5664 &{\color{red}13.2951} & 3.832129  &  9.4501 &{\color{red}13.2822} &  3.47803  &   9.949  &{\color{red}13.4270} \\
 5   &  3.833436   & 9.0834 &{\color{red}12.9168} & 3.56175   & 9.3994 &{\color{red}12.9611} & 3.665086  &  9.283  &{\color{red}12.9480} &  3.31098  &   9.782  &{\color{red}13.0929} \\ 
\end{tabular}
\end{ruledtabular}
\begin{tabbing}
\textcolor{red}{$^{\updownarrow}\ell$ and $m$ are fixed at $0$. \tab[2.5cm] $^{\dagger}n$ and $m$ are fixed at $5$ and $0$ respectively. 
\tab[2.5cm] $^{\ddagger}n$ and $\ell$ both are fixed at $5$. }\\
\textcolor{red}{$^{\P}$Lower bounds of $S_{t}(=S_{\rvec}+S_{\pvec})$ is 6.43418 followed by the Eq. $S_{\rvec}+S_{\pvec}\geq d(1+ \ln \pi)$.}  
\end{tabbing}
\end{table}
\endgroup  

Now we shift our focus on to $S$ in such a potential. First note that, single-particle probability density in an 
arbitrary state in $r$ space may be written as,
\begin{equation}
\rho(\rvec) = {\left|N_{n,\ell}\right|}^2\ e^{-{\xi r}} r^{2{\beta_{\ell}}} \ {\left[L^{2 {\beta_{\ell}} + 1}_{n} 
\left(\xi r \right)\right]}^{2}\ {\left[Y_{\ell,m}\left({\Omega}\right)\right]}^{2}. 
\end{equation}
One can then decompose $S_{\rvec}$ in terms of following five integrals, 
\begin{equation}
S_{\rvec} = -\int {\rho}\left({\rvec}\right) \ln {\rho}\left({\rvec}\right) \mathrm{d}\rvec \\
= - (S_{1} + S_{2} + S_{3} + S_{4} +S_{5})
\end{equation}
where,
\begin{equation}
\begin{aligned}
S_{1} & = \int {\rho}\left({\rvec}\right) \ln {\left|N_{n,\ell}\right|}^2  \mathrm{d}\rvec = \ln {\left|N_{n,\ell}\right|}^2, \\
\textcolor{red}{S_{2}} &= \textcolor{red}{-\int \xi r {\rho}\left({\rvec}\right)   \mathrm{d}\rvec} \ \  
= \textcolor{red}{-|N_{n, \ell}|^{2} \left(\frac{1}{\xi}\right)^{\left(2 \beta_{\ell} +3\right)} \sum_{i=0}^n \left(\begin{array}{c}
 2\\ n-i \end{array}\right)^2 \frac{\Gamma\left(2 \beta_{\ell} +4 + i\right)}{i!}},
\end{aligned}
\end{equation}
\begin{equation}
\begin{aligned}
S_{3} =  & 2 \beta_{\ell} \int {\rho}\left({\rvec}\right) \ln {r} \ \mathrm{d}\rvec 
= -2 \beta_{\ell} |N_{n, \ell}|^{2} \left(\frac{1}{\xi}\right)^{\left(2 \beta_{\ell} +3\right)} \ (\ln {\xi}) 
 \sum_{i=0}^n \left(\begin{array}{c} 1\\ n-i \end{array}\right)^2 \frac{\Gamma\left(2 \beta_{\ell} +3 + i\right)}{i!} + \ \\
& 2 \beta_{\ell} |N_{n, \ell}|^{2} \left(\frac{1}{\xi}\right)^{\left(2 \beta_{\ell} +3\right)} 
 \int t^{\left(2 \beta_{\ell} +2\right)} e^{-t} \left[L^{\left(2 \beta_{\ell} +2\right)}_{n}(t) 
 \right]^2 \ln{t} \ \mathrm{d}t,  \\
S_{4} =& \int {\rho}\left({\rvec}\right) \ln {\left[L^{\left(2 \beta_{\ell} +1\right)}_{n}(\xi r) \right]^2} \mathrm{d} \rvec \\
=& |N_{n, \ell}|^{2} \left(\frac{1}{\xi}\right)^{\left(2 \beta_{\ell} +3\right)} 
\int t^{\left(2 \beta_{\ell} +2\right)} e^{-t} \left[L^{\left(2 \beta_{\ell} +2\right)}_{n}(t)
 \right]^2 \ln {\left[L^{\left(2 \beta_{\ell} +2\right)}_{n}(t)
 \right]^2} \ \mathrm{d}t, 
\end{aligned}
\end{equation}
\begin{equation}
S_{5} = \int \left[Y_{\ell,m}\left({\Omega}\right)\right]^{2} \ln {\left[Y_{\ell,m}\left({\Omega}\right)\right]}^{2} 
\mathrm{d}{\Omega},
\end{equation}
where we have defined $ t = {\xi} r$. Further simplification of integrations given in $S_3$ and $S_4$ requires more detailed knowledge 
about orthogonal polynomials. However, for certain values of $\ell$ ($0,1$) $S_5$ can be computed analytically \cite{yanez94}. In various 
occasions this integral with $\ell=0-9$ has been evaluated numerically \cite{mukherjee18a}. 

\begin{figure}                         
\begin{minipage}[c]{0.3\textwidth}\centering
\includegraphics[scale=0.45]{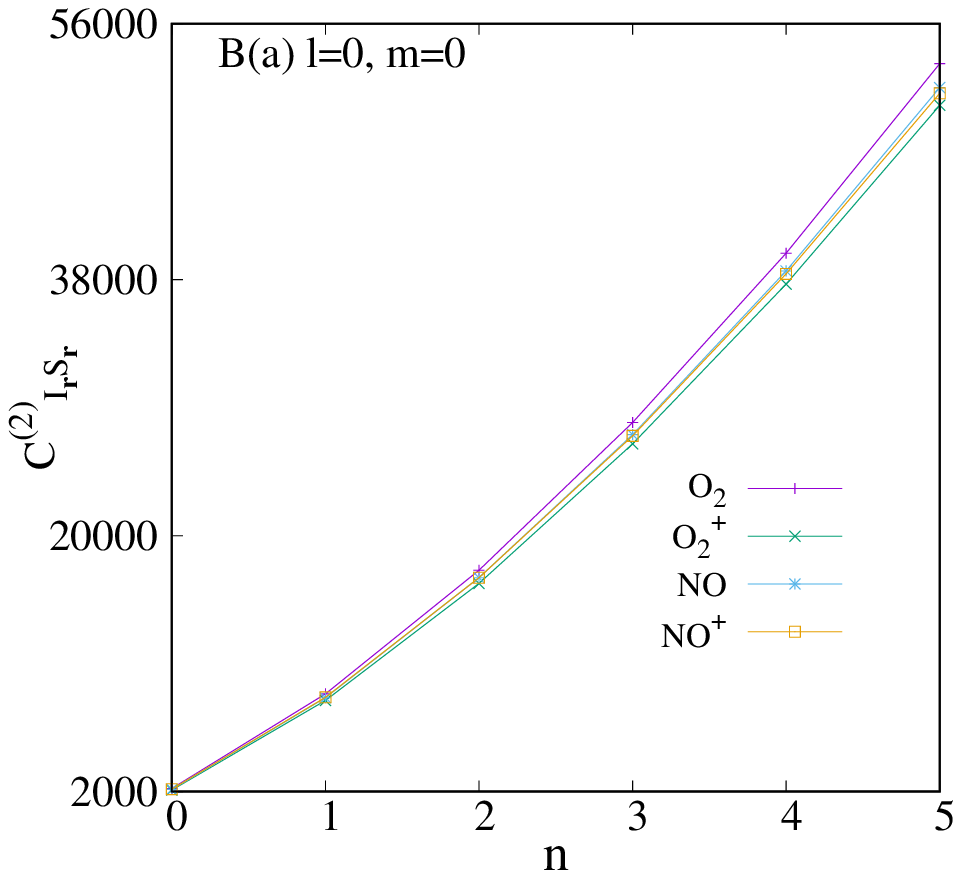}
\end{minipage}%
\begin{minipage}[c]{0.3\textwidth}\centering
\includegraphics[scale=0.45]{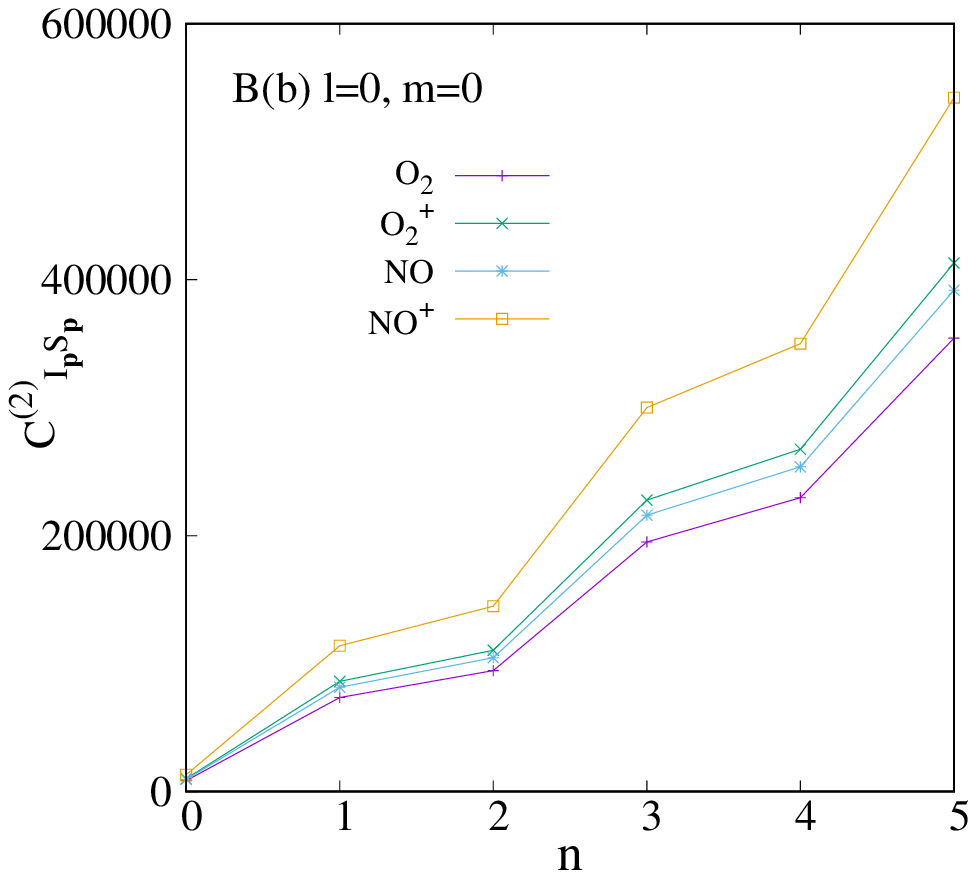}
\end{minipage}%
\vspace{2mm}
\begin{minipage}[c]{0.3\textwidth}\centering
\includegraphics[scale=0.45]{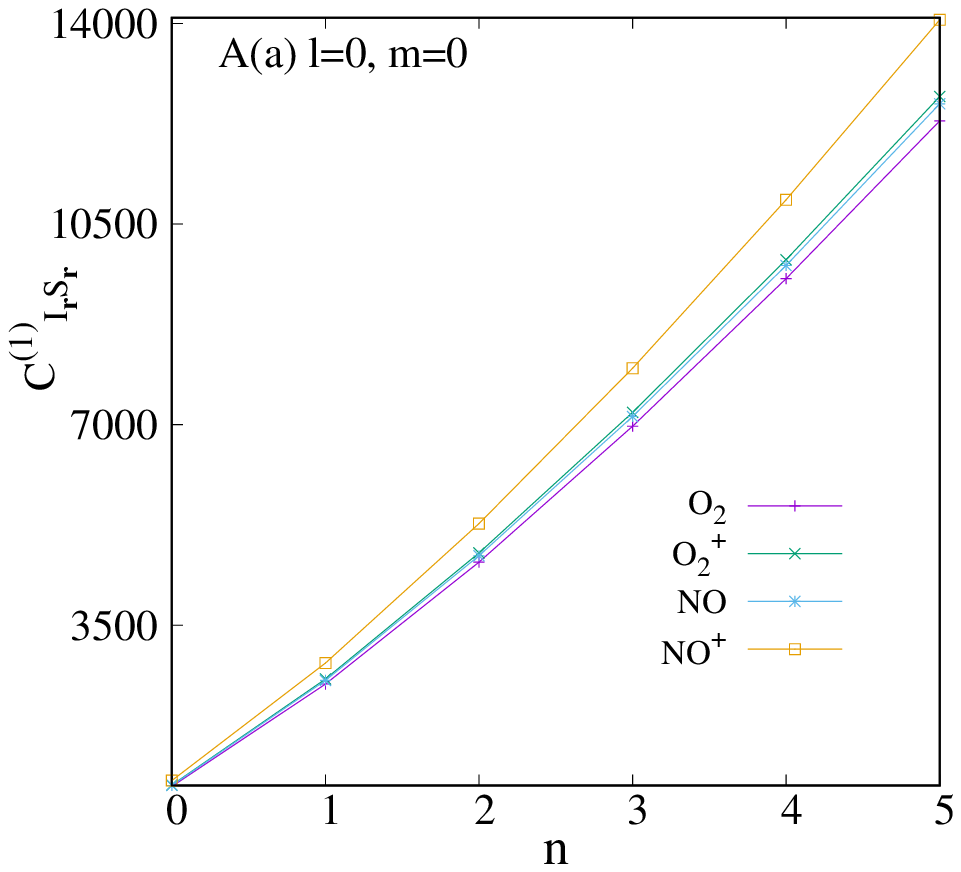}
\end{minipage}%
\begin{minipage}[c]{0.3\textwidth}\centering
\includegraphics[scale=0.45]{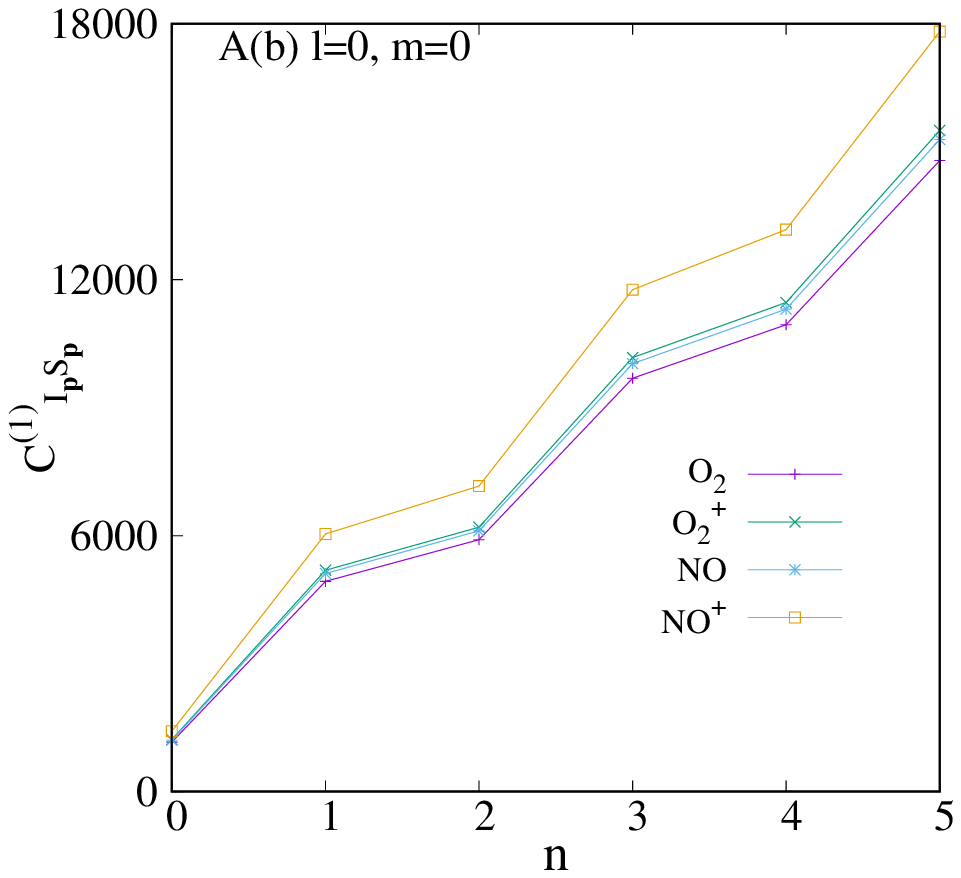}
\end{minipage}%
\caption{Variation of $C_{I_{\rvec}S_{\rvec}}^{(1)},~C_{I_{\pvec}S_{\pvec}}^{(1)}$ (bottom row A) and 
$C_{I_{\rvec}S_{\rvec}}^{(2)},~C_{I_{\pvec}S_{\pvec}}^{(2)}$ (top row B)
in Mie potential with $n$ keeping both $\ell, m$ fixed at zero. Consult text for more details.}
\end{figure}

Next, Table~III reports numerically calculated (from Eq.~(18)) $S_{\rvec}, S_{\pvec}$ for the same set of molecules, as in the previous table.
The same presentation strategy has been adopted in three horizontal blocks to illustrate changes in $n,\ell,m$ respectively
keeping other two fixed at some selected values. From the first block, it is clear that, both measures increase as one goes to higher
$n$, for all molecules considered, implying that states become more and more diffused with the addition of radial nodes in wave 
function. A similar trend is observed in case of $S_{\pvec}$ as well. The middle block indicates a reduction in $S_{\rvec}$  
with progression in $\ell$, whereas $S_{\pvec}$ first falls and rises again passing through a minimum, for a chosen molecule. 
The last block shows that at first they both increase and after reaching maxima they tend to fall off, but the maxima occur 
at different $m$ in two spaces. An interesting point here is that, in contrast to $I_{\rvec}$, $S_{\rvec}$ 
for a neutral molecule assumes higher value than its cationic counterpart. \textcolor{red}{It happens due to the reason that, shifting from 
neutral to cationic species $D_0$ increases and $r_0$ decreases. Therefore the latter has a sharp probability distribution 
(stronger bond) compared to the former. Hence it is evident that an increase in bond strength is indicated by the 
decrease in $S_{\rvec}$. It is worthwhile mentioning that, in all cases, $S_t$ obeys
the lower bound provided in Eq.~(17).} 

\begin{figure}                         
\begin{minipage}[c]{0.3\textwidth}\centering
\includegraphics[scale=0.45]{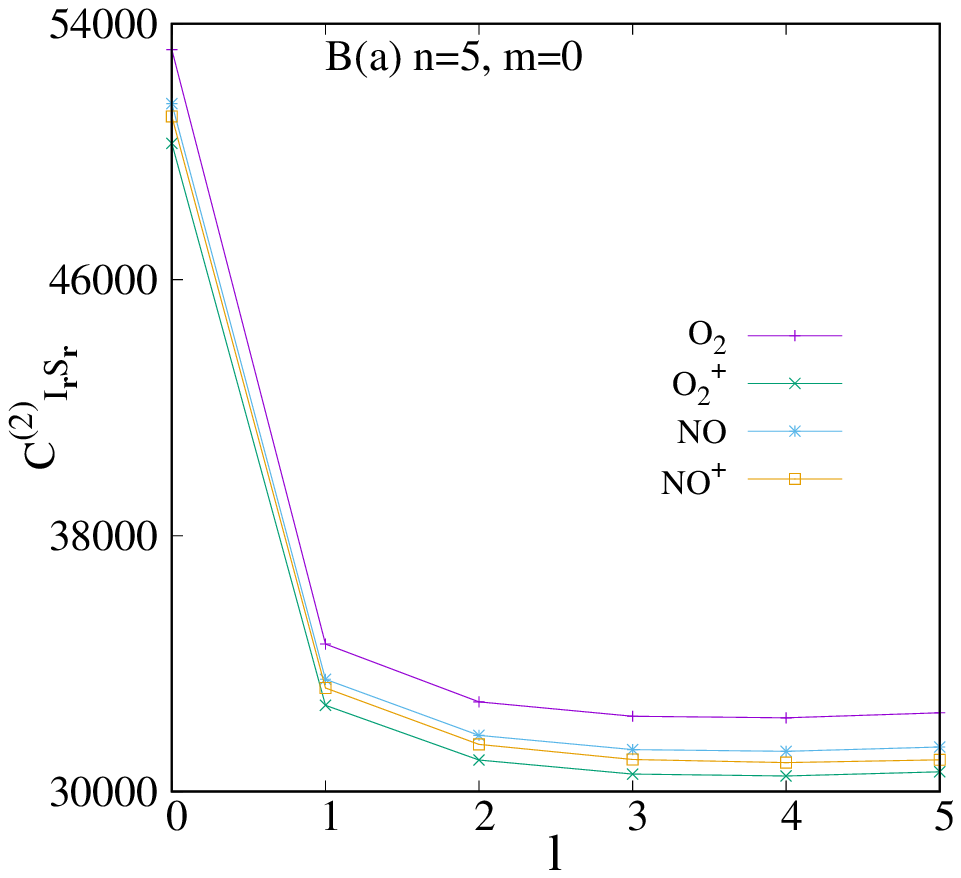}
\end{minipage}%
\begin{minipage}[c]{0.3\textwidth}\centering
\includegraphics[scale=0.45]{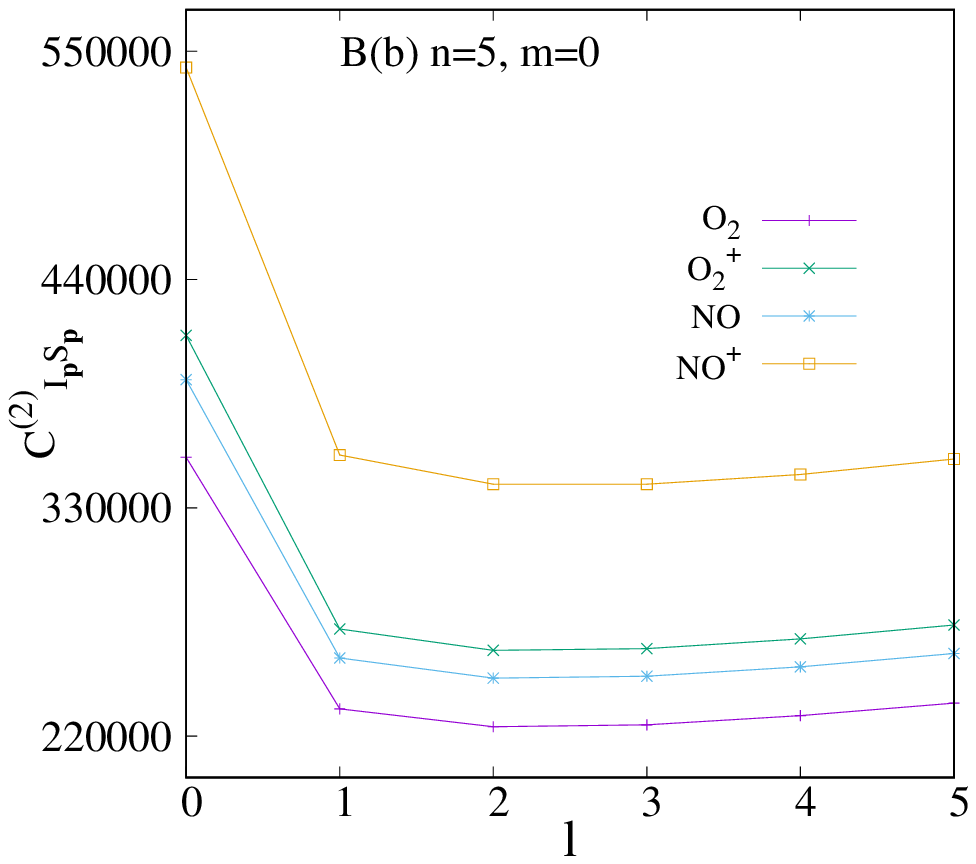}
\end{minipage}%
\vspace{2mm}
\begin{minipage}[c]{0.3\textwidth}\centering
\includegraphics[scale=0.45]{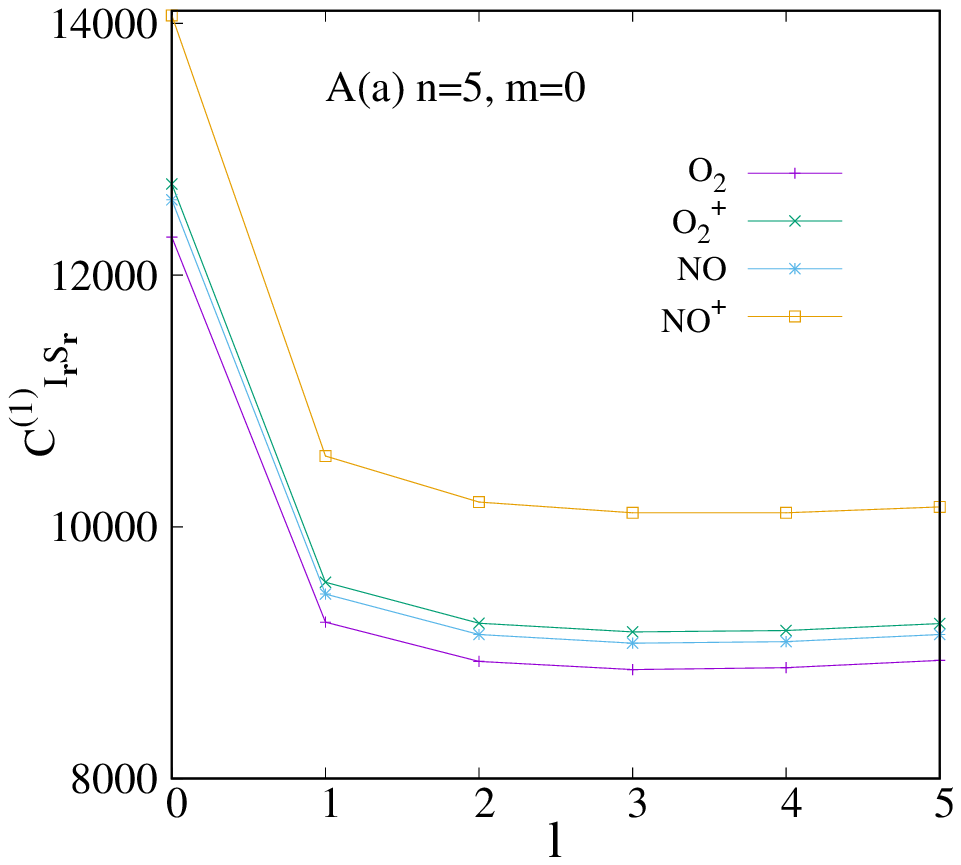}
\end{minipage}%
\begin{minipage}[c]{0.3\textwidth}\centering
\includegraphics[scale=0.45]{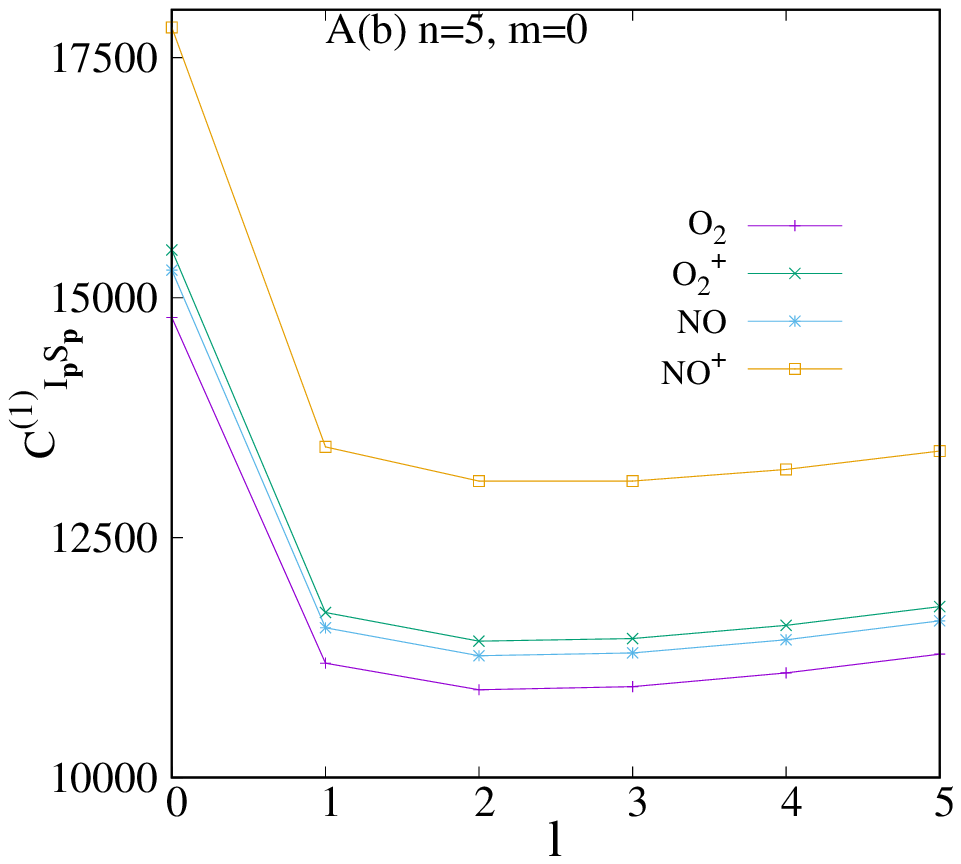}
\end{minipage}%
\caption{Variation of $C_{I_{\rvec}S_{\rvec}}^{(1)},~C_{I_{\pvec}S_{\pvec}}^{(1)}$ (bottom row A)
$C_{I_{\rvec}S_{\rvec}}^{(2)},~C_{I_{\pvec}S_{\pvec}}^{(2)}$ (top row B)
in Mie potential with $\ell$, choosing $n=5, m=0$. Consult text for more details.}
\end{figure}

From the foregoing discussion, we see that, at a specific $\ell,m$, both $I_{\rvec}, S_{\rvec}$ escalate with 
principal quantum number; the former suggesting fluctuation while latter signifying spreading in density distribution. 
The other state indices have also shown analogous changes in these quantities. Thus 
it would be interesting to see their combined effect from a consideration of $C_{IS}$, for which we now proceed.  

In bottom row panels A(a), A(b) of Fig.~1, $C_{I_{r}S_{r}}^{(1)}, C_{I_{p}S_{p}}^{(1)}$ are plotted against
$n$ (from 0-5) keeping $\ell,m$ fixed at 0, while $C_{I_{r}S_{r}}^{(2)}, C_{I_{p}S_{p}}^{(2)}$ are portrayed 
in top panels B(a), B(b) respectively. In all four cases, illustrative calculations are done with the same four molecules 
as earlier. Two segments in first column reveal that, both 
$C_{I_{r}S_{r}}^{(1)}, C_{I_{r}S_{r}}^{(2)}$ grow as $n$ progresses. This can be understood from the data presented 
in Tables~II, III, where we found that both $I, S$ increase with $n$ (for fixed $\ell,m$). Since this implies a 
rise in number of nodes, it appears that addition of nodes facilitates the system to approach towards \textcolor{red}{disorder.
 This pattern resembles the Fig.~1(I) in reference \cite{shiner99}; the monotonically increasing function of $n$ indicating a rise 
in disorder}. 
\textcolor{red}{Similarly for the $p$ space as well, $C_{IS}$ increases with $n$ for both $b$'s but few humps can be noticed.} 
At a given $n$, $C_{I_{r}S_{r}}^{(1)}$ increases from neutral to ionic species; however, this almost coincide for two 
iso-electronic molecules (O$_{2}^{+}$, NO). 
{\color{red} Shifting from neutral to cationic system $r_0$ decreases indicating 
an increase in bond strength with the rise in $I_{\rvec} (S_{\pvec})$ and a decrease in $S_{\rvec} (I_{\pvec})$. For $b=\frac{2}{3}$, 
$I$ dominates over $S$, causing a rise in complexity measure in $r$ space with increasing bond strength. 
Similar ordering of the neutral and ionic system for a particular state with respect to this complexity measure in $p$ space 
is observed. In contrast to the former, for $b=1$, $C_{I_{r}S_{r}}^{(2)}$ for neutral molecule possesses higher 
value than its cationic analogue. This indicates the domination of $S$ over $I$ in the measured quantity while shifting from 
neutral to cationic equivalent at a particular state.
 However, in $p$ space, $C^{(2)}_{IS}$ decreases from cation to neutral molecule which is opposite to its conjugate $r$-space 
ordering. This pattern was lost for $b=\frac{2}{3}$. From this point of view, it seems that $b=1$ possibly qualifies to be a 
more appropriate descriptor for the systems under investigation.} 

Similarly, panels A(a), A(b) of the bottom row of Fig.~2 illustrate behaviour of $C_{I_{r}S_{r}}^{(1)}$, $C_{I_{p}S_{p}}^{(1)}$ 
with changes in $\ell$, while keeping other two quantum numbers $n,m$ fixed at $5, 0$ respectively. Analogously, 
$C_{I_{r}S_{r}}^{(2)}$, $C_{I_{p}S_{p}}^{(2)}$ are pictorially represented in top panels B(a), B(b). Here, both 
$C_{I_{r}S_{r}}^{(1)}, C_{I_{r}S_{r}}^{(2)}$ decline with an increase in $\ell$ having a fixed number of nodes, which acts as 
an indicator of the system approaching towards disorder. \textcolor{red}{This may be compared to the Fig.~1(III) 
in reference \cite{shiner99} which depicted complexity as a measure of order as decreases with increasing disorder of the
 system. In $p$ space, both complexity 
measures first decay, and after reaching a minimal value slightly go up with growth in $\ell$. 
As noticed in Fig.~1, this figure also supports the fact that $b=1$ possibly characterizes the system in a more appropriate 
manner. One notices that, 
while $C_{I_{r}S_{r}}^{(2)}$ and $C_{I_{p}S_{p}}^{(2)}$ show reverse ordering for a particular state in the ionic and neutral 
species, for $C_{I_{r}S_{r}}^{(1)}, C_{I_{p}S_{p}}^{(1)}$ no such reversal occurs.}

\begin{figure}                         
\begin{minipage}[c]{0.3\textwidth}\centering
\includegraphics[scale=0.45]{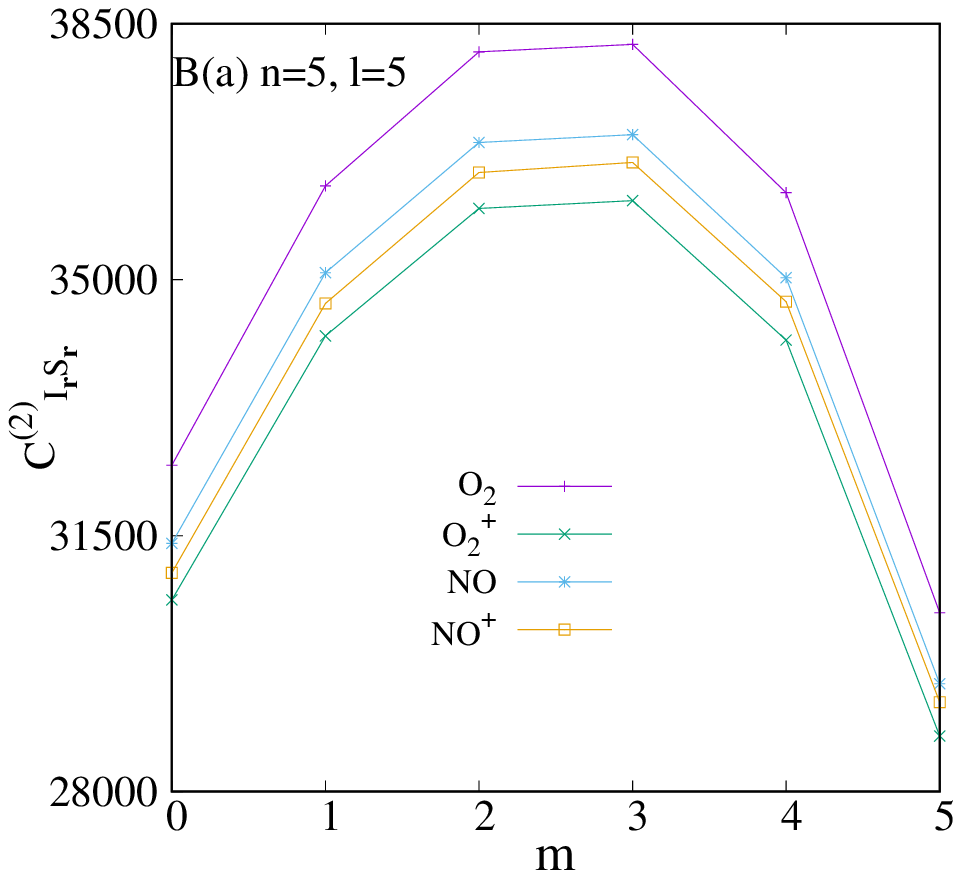}
\end{minipage}%
\begin{minipage}[c]{0.3\textwidth}\centering
\includegraphics[scale=0.45]{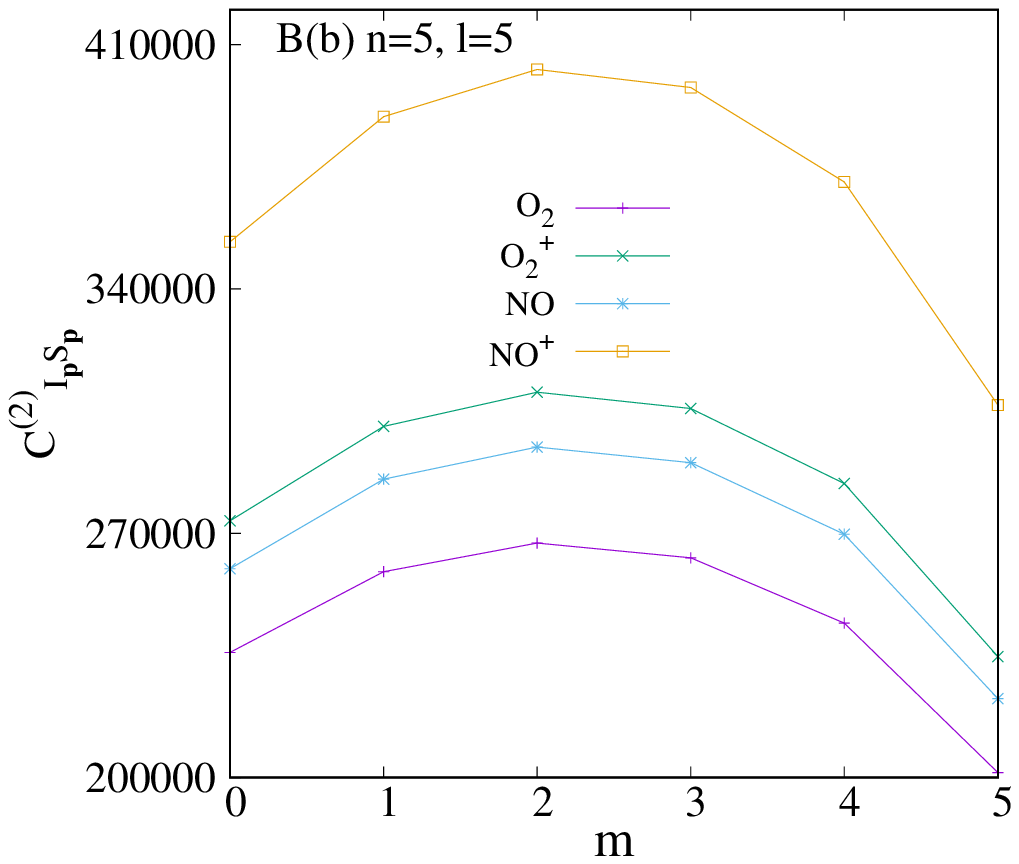}
\end{minipage}%
\vspace{2mm}
\begin{minipage}[c]{0.3\textwidth}\centering
\includegraphics[scale=0.45]{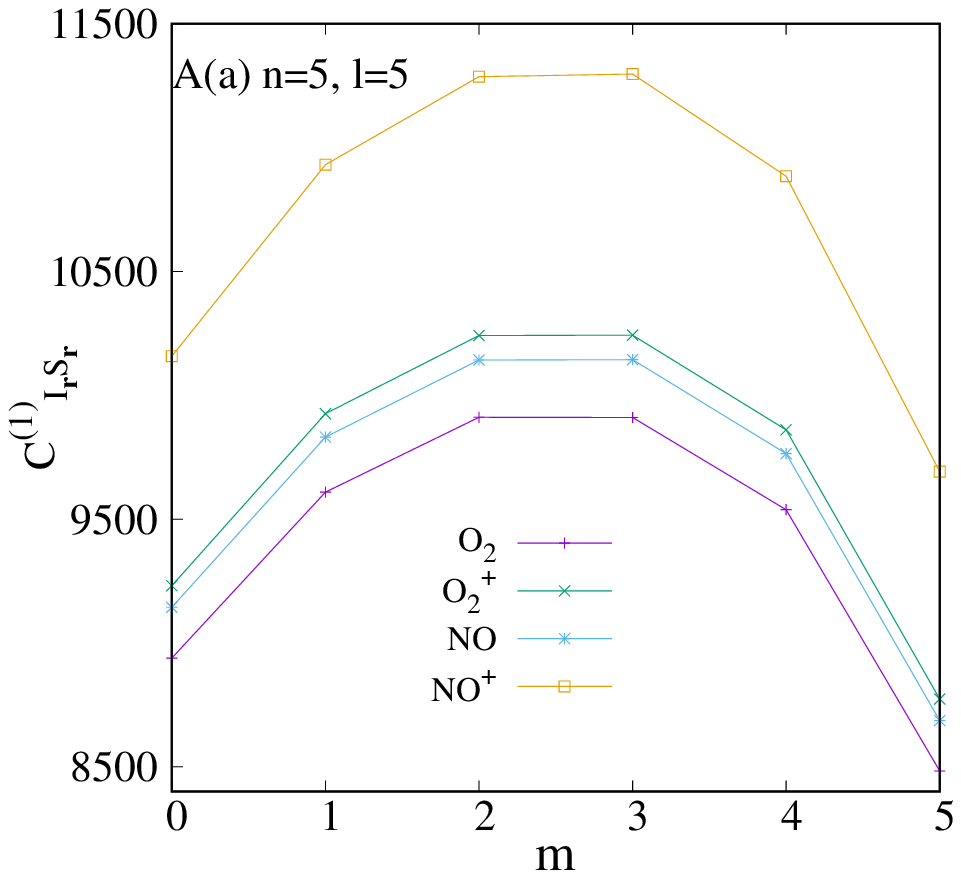}
\end{minipage}%
\begin{minipage}[c]{0.3\textwidth}\centering
\includegraphics[scale=0.45]{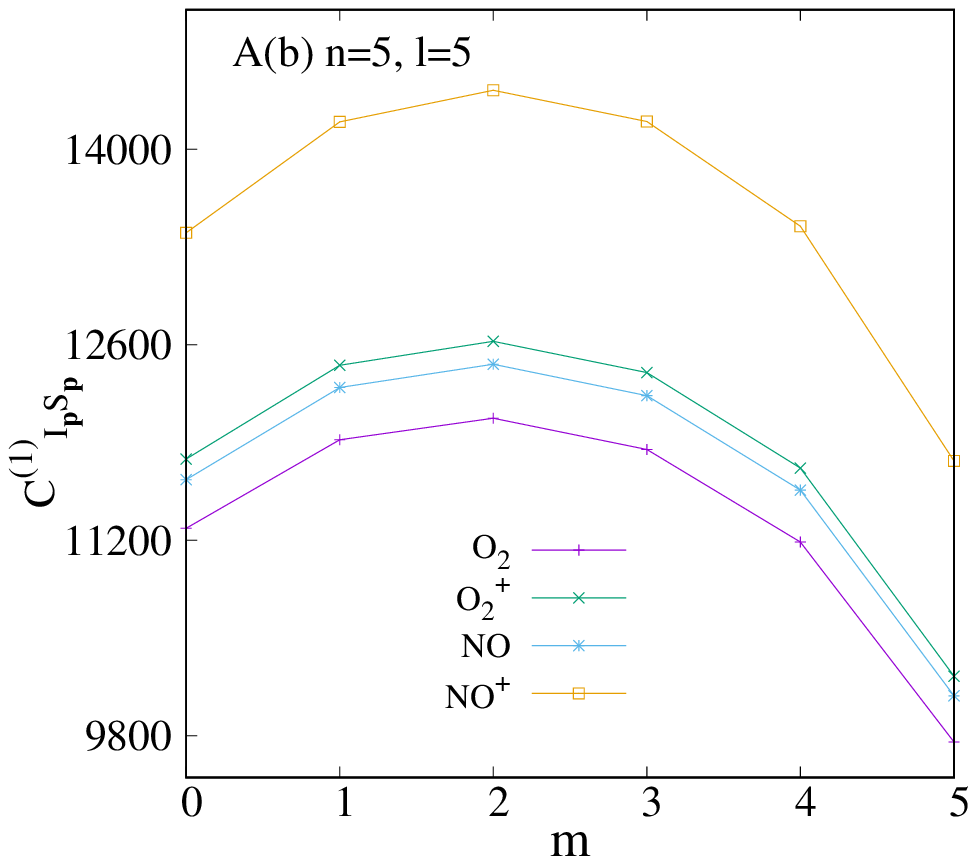}
\end{minipage}%
\caption{Variation of $C_{I_{\rvec}S_{\rvec}}^{(1)},~C_{I_{\pvec}S_{\pvec}}^{(1)}$ (bottom row A) and 
$C_{I_{\rvec}S_{\rvec}}^{(2)},~C_{I_{\pvec}S_{\pvec}}^{(2)}$ (top row B)
in Mie potential with $m$ choosing $n=5, \ell=5$. Consult text for more details.}
\end{figure}

Finally, Fig.~3 illustrates the nature of $C_{I_{r}S_{r}}^{(1)}$, $C_{I_{p}S_{p}}^{(1)}$ and $C_{I_{r}S_{r}}^{(2)}$, 
$C_{I_{p}S_{p}}^{(2)}$ with changes in $m$ in panels \{A(a), A(b)\} and \{B(a), B(b)\} respectively, keeping $n,\ell$ both 
fixed at $5$. \textcolor{red}{In both spaces, as $m$ rises, $C_{IS}$ for each molecule gradually increase and after 
passing through a maximum, falls off. This feature of complexity is a characteristic of the quantum system 
lying between order and disorder. A similar trend is depicted in Fig.~1(II) of reference \cite{shiner99}.} 
\textcolor{red}{ Moreover, like earlier two cases, here also $C_{I_{\rvec}S_{\rvec}}^{(2)}$ 
possesses larger values for neutral species than ions, but for $C_{I_{\pvec}S_{\pvec}}^{(2)}$ reverse trend is observed.}

\section{Concluding Remarks}
Information theoretical measures such as $I,~S,~C_{IS}$ are pursued on some diatomic molecules with generalized 
Kratzer type potential, in conjugate $r$ and $p$ spaces. These are analyzed in a systematic manner, for both ground and excited 
states. Exact analytical expressions of $I_{\rvec}$ are provided for any given state having arbitrary values for quantum numbers 
$n,\ell,m$; for $I_{\pvec}$ these are derived for only $m=0$ states. Accurate numerical results are presented for four molecules 
including two cations. Also, an attempt has been made to derive expressions for $S_{\rvec}$; however, in this case, only partial 
analytical expressions have been established in terms of certain entropic integrals. Interesting patterns are observed in the 
behaviour of $I$ and 
$S$ in the cationic and neutral molecules. \textcolor{red}{Though both $S$ and $I$ increase with a rise in $n$, 
the former suggests growth in delocalization while the latter indicates an escalation in fluctuation. With the increase in $\ell$, 
the variation of both the quantities $I$ and $S$ are consistent in capturing localization. Depending on the choice of 
quantum numbers ($n,~\ell,~m$), $C_{IS}$ seems to approach all the three categories mentioned in the reference \cite{shiner99}. 
Furthermore, the investigation of $C_{IS}$ complexity establishes that $b=1$ characterizes the system in a more 
proper manner than $b=\frac{2}{3}$ for the molecules studied here}. It would be interesting to explore other 
measures like R\'enyi entropy, Onicescu energy, Tsallis entropy, as well as other complexity measures. 

\section{Acknowledgement}
{\color{red} SM is grateful to IISER Kolkata for Junior Research Fellowship (JRF). NM acknowledges SERB-INDIA for National post-doctoral 
fellowship (sanction order: PDF/2016/000014/CS). Financial support from DST SERB New Delhi, India (sanction order: EMR/2014/000838) 
is gratefully acknowledged. Constructive comments from two anonymous 
referees have been very helpful. Prof.~A.~K.~Nanda is thanked for many enlightening discussion. }

\end{document}